\documentclass[twocolumn,preprintnumbers,amsmath,amssymb,pra]{revtex4}
\usepackage{graphicx}
\usepackage{dcolumn}
\usepackage{bm}
\usepackage{times}
\usepackage[colorlinks,citecolor=blue,linkcolor=red]{hyperref}
\usepackage{color}
\usepackage{comment}

\begin{document}

\title{Nonequilibrium thermal transport and photon squeezing in a quadratic qubit-resonator system}

\author{Chen Wang$^{1,}$}
\author{Hua Chen$^{1}$}\email{hwachanphy@zjnu.edu.cn}
\author{Jie-Qiao Liao$^{2,}$}\email{jqliao@hunnu.edu.cn}
\address{
$^{1}$Department of Physics, Zhejiang Normal University, Jinhua 321004, Zhejiang , P. R. China\\
$^{2}$Key Laboratory of Low-Dimensional Quantum Structures and Quantum Control of Ministry of Education, Key Laboratory for Matter Microstructure and Function of Hunan Province,
Department of Physics and Synergetic Innovation Center for Quantum Effects and Applications, Hunan Normal University, Changsha 410081, China
}

\date{\today}

\begin{abstract}
We investigate steady-state thermal transport and photon statistics in a nonequilibrium hybrid quantum system,
in which a qubit is longitudinally and quadratically coupled to an optical resonator.
Our calculations are conducted with the method of the quantum dressed master equation combined with full counting statistics.
The effect of negative differential thermal conductance is unravelled at finite temperature bias,
which stems from the suppression of cyclic heat transitions and large mismatch of two squeezed photon modes at weak and strong qubit-resonator hybridizations, respectively.
The giant thermal rectification is also exhibited at large temperature bias.
It is found that the
intrinsically asymmetric structure of the hybrid system and negative differential thermal conductance
show  the cooperative contribution.
Noise power and skewness, as typical current fluctuations, exhibit global maximum with strong hybridization at small and large temperature bias limits, respectively.
Moreover, the effect of photon quadrature squeezing is discovered in the strong hybridization and low-temperature regime,
which shows asymmetric response to two bath temperatures.
These results would provide some insight to thermal functional design and photon manipulation in qubit-resonator hybrid quantum systems.
\end{abstract}

\maketitle

\section{Introduction}
The efficient control of energy transport and information processing is of fundamental importance
in low dimensional nano-structures~\cite{gchen2005book},
which spurs the advance of interdisciplinary frontiers,
ranging from quantum thermodynamics and thermal machines~\cite{gbenenti2017pr,mesposito2019prl}, quantum electrodynamics (QED)~\cite{aronzani2018np,jsenior2020cp,dwang2019np} to topological quantum optics~\cite{hxu2016nature,sbarik2018sci,ywang2020pr}.
Of particular interest is the nonequilibrium thermal transport in open small quantum systems,
where the degrees of freedom of the target system are much fewer than those of the external baths~\cite{uweiss2012book}.
Accordingly, the microscopic modeling of the system-bath interaction is crucial for
theoretical characterization of the heat flow.
Moreover, the strong system-bath coupling significantly fertilizes the physical picture of thermal transport~\cite{dsegal2006prb,dsegal2011prb,ksaito2013prl,mcarrega2016prl,wdou2018prb,anazir2020prl}.
Hence, it is intriguing to understand the underlying mechanism of the nonequilibrium thermal transport in small quantum systems.

One of the representative small quantum systems to describe nonequilibrium thermal transport probably is the
nonequilibrium spin-boson model (SBM)~\cite{aleggett1987rmp}, which is composed by a two-level system interacting with two individually bosonic thermal baths.
The heat flow in the nonequilibrium SBM has been extensively investigated from weak to strong system-bath couplings by various analytical approaches,
including the Redfield equation~\cite{dsegal2005prl}, noninteracting blip approximation~\cite{dsegal2006prb,dsegal2011prb},
nonequilibrium polaron-transformed Redfield equation~\cite{cwang2015sr,cwang2017pra,xfcao2021prb},
nonequilibrium green function~\cite{yy2014epl,jjliu2017pre}, and reaction coordinate mapping~\cite{anazir2020prl}.
While the rise of quantum engineering spawns another typical small quantum system platform, the hybrid quantum system (HQS)~\cite{mwallquist2009ps,gkurizki2015pnas,aclerk2020np}.
For example, the circuit-QED system~\cite{akockum2019nrp,pdiaz2019rmp} and optomechanical system~\cite{maspelmeyer2014rmp},
they enable the strong hybridization of the artificial qubit and mechanical phonon with optical photon, respectively.
Interestingly, the reaction coordinate mapping tries to theoretically establish the bridge between the SBM and the coupled qubit-resonator  system~\cite{anazir2014pra,anazir2020prl,anazir2016jcp,dnewman2017pre,srestrepo2018njp,cmc2019jcp}.
Recently, two-photon based quadratic interaction in HQSs, e.g., two-photon Rabi model, has attracted increasing attention,
which has practical correspondence in the circuit-QED platform~\cite{sfelicetti2018pra1,sfelicetti2018pra2}, and
theoretically leads to spectral collapse~\cite{qhchen2012pra,lduan2016jpa}, quantum phase transition~\cite{lgarbe2017pra,xchen2018pra,scui2020pra},
{ultrafast charging quantum battery~\cite{ac2020prb,ad2021entropy} and enhancement of superradiant spontaneous emission~\cite{nparxiv2020}.
}
Moreover, such nonlinear interactions play a crucial role in modeling the electric current based on the quantum dot-vibration hybrid model~\cite{jjliu2020nl}.

From the aspect of quantum thermal transport, increasing works are carried out in HQSs.
Specially, chiral thermal management~\cite{aseif2017nc,zdenis2020prl} and Berry-phase-like heat pump~\cite{wjnie2020pra} are proposed in optomechanical systems, which result in non-reciprocal heat flow.
Recently, Wu \emph{et al.}~\cite{yang2020nc} experimentally analyzed the phonon heat transport in a mechanical resonator dimer mediated by one optical cavity mode,
which directly measured heat current and proved thermodynamics uncertainty relation.
While for the coupled qubit-resonator system, typical thermal functional devices,
e.g., quantum thermal transistor and quantum rectifier, have been proposed by linearly hybridizing the qubit to the optical resonator
with the longitudinal and transverse forms~\cite{mmajland2020prb,cwang2021cpl,cwang2021cpb,yjzhao2015pra,xwang2016pra,xwang2017pra,sricher2016prb,nlambert2018prb}.
The novel two-peak feature of linear thermal conductance by tuning the bath temperature was unravelled, which cannot be simply observed in the SBM~\cite{tkato2020arxiv}.
Meanwhile, the heat flow was practically measured in the circuit-QED setup with weak qubit-photon hybridization~\cite{aronzani2018np,jsenior2020cp}.

In this paper, we study the nonequilibrium thermal transport and photon squeezing in a {longitudinal and quadratic coupling} qubit-resonator system.
Concretely,
for the steady-state heat current, we clearly unravel the effect of negative differential thermal conductance (NDTC) in a wide qubit-photon interaction regime by tuning the bath temperature bias.
We also find the giant thermal rectification at strong qubit-photon hybridization.
For the heat current fluctuations, we analyze the noise power and skewness, which show global maximum with strong hybridization strength at small and large temperature biases, respectively.
While for the photon squeezing, we find that strong qubit-photon hybridization and low bath temperatures is helpful to exhibit the photon squeezing.
Moreover, the photon squeezing shows asymmetric response to two bath temperatures.

The rest of this paper is organized as follows.
In Sec. II, we introduce the coupled qubit-resonator model with quadratic interaction,
derive the dressed master equation (DME) to obtain the steady state,
and combine DME with full counting statistics (FCS) to obtain the heat current fluctuations.
In Sec. III, we first study thermal functionalities, i.e., the NDTC and heat amplification, by  modulating both the qubit-photon hybridization strength and the bath temperature bias.
We also investigate the steady-state noise power and skewness as the representative current fluctuations.
Finally, we analyze the effect of photon squeezing both at equilibrium and out of equilibrium.
In Sec. IV, we present a conclusion.


\section{Model and method}
\subsection{Qubit-resonator hybrid system}
\begin{figure}[tbp]
\includegraphics[scale=0.35]{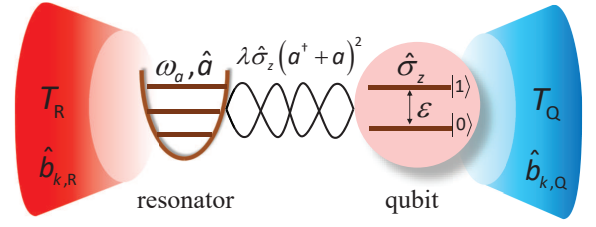}
\caption{(Color online)
(a) Schematic of the nonequilibrium coupled qubit-resonator system.
The harmonics and  two-level system denote the optical resonator and qubit, respectively.
The pair of double black crossing curves shows the quadratic qubit-photon interaction;
the red and blue panels represent the bosonic thermal baths, characterized as the temperatures $T_\textrm{R}$ and $T_\textrm{Q}$, respectively.
}~\label{fig1}
\end{figure}

The nonequilibrium hybrid quantum system under consideration in Fig.~\ref{fig1} is composed by an optical resonator quadratically interacting with a qubit, each coupled to a bosonic thermal bath.
The Hamiltonian of the coupled qubit-resonator system reads~($\hbar=1$)
\begin{eqnarray}~\label{hs0}
\hat{H}_\textrm{S}=\omega_a\hat{a}^\dag\hat{a}+\frac{\varepsilon}{2}\hat{\sigma}_z+\lambda\hat{\sigma}_z(\hat{a}^\dag+\hat{a})^2,
\end{eqnarray}
where $\hat{a}^\dag~(\hat{a})$ denotes the creation  (annihilation) operator of the optical resonator with the resonant frequency $\omega_a$,
$\sigma_{x}=|1{\rangle}{\langle}0|+|0{\rangle}{\langle}1|$ and $\sigma_{z}=|1{\rangle}{\langle}1|-|0{\rangle}{\langle}0|$ are the Pauli operators of the qubit composed by two states $|1{\rangle}$ and $|0{\rangle}$,
and $\lambda$ is the qubit-photon interaction strength.
Thereafter, we set $\omega_a=1$ without losing any generality.
The $u$th thermal bath is described as noninteracting bosons $\hat{H}^u_\textrm{B}=\sum_k\omega_{k,u}\hat{b}^\dag_{k,u}\hat{b}_{k,u}$,
where $\hat{b}^\dag_{k,u}~(\hat{b}_{k,u})$ is the creation(annihilation) operator of the bosonic mode with the momentum $k$ and the frequency $\omega_{k,u}$.
The interactions of the photon and qubit with the corresponding thermal baths are given by
\begin{subequations}
\begin{align}
\hat{V}_{\textrm{R}}=&(\hat{a}^\dag+\hat{a})\sum_k(g_{k,\textrm{R}}\hat{b}^\dag_{k,\textrm{R}}+g^*_{k,\textrm{R}}\hat{b}_{k,\textrm{R}}),~\label{vr}\\
\hat{V}_{\textrm{Q}}=&\hat{\sigma}_x\sum_k(g_{k,\textrm{Q}}\hat{b}^\dag_{k,\textrm{Q}}+g^*_{k,\textrm{Q}}\hat{b}_{k,\textrm{Q}})~\label{vq},
\end{align}
\end{subequations}
with $g_{k,\textrm{R} (\textrm{Q})}$ being the photon (qubit)-boson interaction strengthes.
Based on the above description, it is known that the Hamiltonian of the whole system, which both include the hybrid system and the environment,
is described by
\begin{eqnarray}
\hat{H}=\hat{H}_\textrm{S}+\sum_{u=\textrm{R},\textrm{Q}}(\hat{H}^u_\textrm{B}+\hat{V}_u).
\end{eqnarray}

It is interesting to point out that the system Hamiltonian given in Eq.~(\ref{hs0}) can be exactly solved.
Under the basis states $|1{\rangle}$ and $|0{\rangle}$  of the qubit, the Hamiltonian in Eq.~(\ref{hs0}) can be decomposed as
$\hat{H}_\textrm{S}=\sum_{\sigma=1,0}\hat{H}^\sigma_s|\sigma{\rangle}{\langle}\sigma|$,
with
$\hat{H}^\sigma_\textrm{S}=(\omega_a+2\lambda_\sigma)\hat{a}^\dag\hat{a}+\lambda_\sigma(\hat{a}^{\dag2}+\hat{a}^2)
+({\varepsilon_\sigma}/{2}+\lambda_\sigma)$,
where the renormalized quantities are given by $\eta_\sigma=\sqrt{1-({2\lambda}/{\omega_\sigma})^2}$,
$\omega_\sigma=\omega_a+2\lambda_\sigma$,
$\varepsilon_{\sigma}=(-1)^{\sigma+1}\varepsilon$,
and
$\lambda_{\sigma}=(-1)^{\sigma+1}\lambda$.
Then, by applying the transformation operator $\hat{U}=\exp\{{-\sum_{\sigma=1,0}{\alpha_\sigma}/{2}|\sigma{\rangle}{\langle}\sigma|{\otimes}(\hat{a}^2-\hat{a}^{\dag2})}\}$
to $\hat{H}_\textrm{S}$ with
\begin{eqnarray}
\tanh\alpha_\sigma=(-1)^{\sigma+1}\sqrt{\frac{1-\eta_\sigma}{1+\eta_\sigma}},
\end{eqnarray}
we find that the transformed Hamiltonian $\hat{U}\hat{H}_\textrm{S}\hat{U}^\dag$ is fully diagonalized. From the aspect of inverse design,
we can express $\hat{H}_\textrm{S}$ in a concise way as
\begin{eqnarray}~\label{hs1}
\hat{H}_\textrm{S}&=&\sum_{\sigma=0,1}[\eta_\sigma\omega_\sigma\hat{A}^\dag_\sigma\hat{A}_\sigma
+({\eta_\sigma-1})\omega_a/{2}+{\varepsilon_\sigma}/{2}+\lambda_\sigma]\nonumber\\
&&{\otimes}|\sigma{\rangle}{\langle}\sigma|,
\end{eqnarray}
where the new bosonic  operators are specified as
$\hat{A}^\dag_\sigma=f_\sigma\hat{a}^\dag+g_\sigma\hat{a}$
and
$\hat{A}_\sigma=f_\sigma\hat{a}+g_\sigma\hat{a}^\dag$,
fulfilling the commutation relation $[\hat{A}_\sigma,\hat{A}^\dag_\sigma]=1$,
and the coefficients are given by
\begin{eqnarray}~\label{fg1}
f_\sigma=\sqrt{\frac{1+\eta_\sigma}{2\eta_\sigma}},\hspace{0.5cm}g_\sigma=(-1)^{\sigma+1}\sqrt{\frac{1-\eta_\sigma}{2\eta_\sigma}}.
\end{eqnarray}
Accordingly, the eigen-equation of $\hat{H}_\textrm{S}$ is given by
$\hat{H}_\textrm{S}|\psi^\sigma_n{\rangle}=E^\sigma_{n}|\psi^\sigma_n{\rangle}$.
The $n$th eigenvalue is given by
$E^\sigma_{n}=\eta_\sigma\omega_\sigma{n}+[{(\eta_\sigma-1)}\omega_a/{2}+{\varepsilon_\sigma}/{2}+\lambda_\sigma]$,
and the corresponding eigenstate is
$|\psi^\sigma_n{\rangle}=(1/{\sqrt{n!}}){\hat{A}^{{\dag}n}_{{\sigma}}}|\psi^\sigma_0{\rangle}$,
where the squeezed vacuum state is given by
\begin{eqnarray}
|\psi^\sigma_0{\rangle}=|\sigma{\rangle}{\otimes}(1-|\phi_\sigma|^2)^{1/4}\sum^{\infty}_{n=0}\frac{\phi^n_{\sigma}}{2^nn!}(\hat{a}^\dag)^{2n}|0{\rangle}_a,
\end{eqnarray}
with the ratio $\phi_\sigma=-g_\sigma/f_\sigma$~($|\phi_\sigma|<1$) and the bare vacuum state $\hat{a}|0{\rangle}_a=0$.
In the squeezing basis, the frequencies of two squeezing bosonic modes are renormlized by the factor $\eta_\sigma$
and displaced by $\pm2\lambda_\sigma$, respectively.
Hence, in the strongly coupled qubit-photon case, it is straightforward to find that $\eta_1\omega_1{\gg}\eta_0\omega_0$
and the hybridization strength is bounded by $\lambda<0.25\omega_a$.

{
Recently, quantum heat transport in qubit-resonator hybrid devices have been theoretically proposed~\cite{tkato2020arxiv,mmajland2020prb,cwang2021cpl,cwang2021cpb} and experimentally established~\cite{aronzani2018np,jsenior2020cp} based on the circuit-QED setups,
where the superconducting qubit can both longitudinally and transversely interact with the photon resonator~\cite{yjzhao2015pra,xwang2016pra,xwang2017pra,sricher2016prb,nlambert2018prb}.
While for the present model Eq.~(\ref{hs0}) with quadratic interaction, it could also be realized in the superconducting quantum circuit, e.g., no dc current biasing the SQUID~\cite{sfelicetti2018pra1}, where the qubit shows longitudinal quadratic coupling with the resonator.
Moreover, the bosonic thermal baths are simulated by or the LC circuit coupled to a resistor~\cite{mmajland2020prb,aaclerk2010rmp}.
Hence, we may be able to analyze heat energy in the circuit-QED system under the temperature bias.
}

\subsection{Quantum master equation}

When considering  weak system-bath interactions, we are able to separately perturb  the photon-bath and qubit-bath interactions.
Under the Born-Markov approximation, the total density operator can be  decomposed as
$\hat{\rho}_{\textrm{tot}}{\approx}\hat{\rho}_\textrm{S}{\otimes}\hat{\rho}_\textrm{B}$,
where $\hat{\rho}_\textrm{S}$ is the density operator of the reduced qubit-resonator  system,
and $\hat{\rho}_\textrm{B}=\frac{1}{Z}\Pi_{u=\textrm{R},\textrm{Q}}\exp[-\hat{H}^u_\textrm{B}/(k_BT_u)]$ is the density operator of thermal baths at equilibrium,
with $Z=\textrm{Tr}_\textrm{B}\{\Pi_{u=\textrm{R},\textrm{Q}}\exp[-\hat{H}^u_\textrm{B}/(k_BT_u)]\}$ the partition function,
$T_u$ the temperature of the $u$th bath, and $k_B$ is the Boltzmann constant ($k_B$ is set to $1$ for convenience).
Then, the quantum master equation is obtained as {(see Appendix)}
\begin{eqnarray}~\label{qme0}
\frac{d}{dt}\hat{\rho}_\textrm{S}&=&i[\hat{\rho}_\textrm{S},\hat{H}_\textrm{S}]
+\frac{1}{2}\sum_{\omega,\omega^\prime;u=\textrm{R},\textrm{Q}}
\{\kappa_u(\omega^\prime)[\hat{P}_u(\omega^\prime)\hat{\rho}_s,\hat{P}_u(\omega)]\nonumber\\
&&+\mathrm{h.c.}\},
\end{eqnarray}
where the rate is given by $\kappa_u(\omega)=\gamma_u(\omega)n_u(\omega)$,
with $\gamma_u(\omega)=2\pi\sum_k|g_{k,u}|^2\delta(\omega-\omega_k)$ the spectral function
and $n_u(\omega)=1/[\exp(\omega_k/T_u)-1]$ the Bose-Einstein distribution function.
The spectral function is selected to be super-Ohmic case
$\gamma_u(\omega)=\pi\alpha_u({\omega^3}/{\omega^2_c})e^{-\omega/\omega_c}\theta(\omega)$,
where $\alpha_u$ is the coupling strength, $\omega_c$ is the cut-off frequency,
and the heaviside step function is $\theta(x)=1$ for $x{\ge}0$ and $\theta(x)=0$ for $x<0$.
The projecting operator of the resonator originates from
$[\hat{a}^\dag(-\tau)+\hat{a}(-\tau)]=\sum_{\omega}\hat{P}_{\textrm{R}}(\omega)e^{-i\omega\tau}$,
which is specified as
$\hat{P}_{\textrm{R}}(\eta_\sigma\omega_\sigma)=\sum_{\sigma}(f_\sigma-g_\sigma)|\sigma{\rangle}{\langle}\sigma|\hat{A}^\dag_\sigma$
and
$\hat{P}_{\textrm{R}}(-\eta_\sigma\omega_\sigma)=\hat{P}^\dag_{\textrm{R}}(\eta_\sigma\omega_\sigma)$.
While the projecting operator of the qubit,
based on the relation
$\hat{\sigma}_x(-\tau)=\sum_{\omega}\hat{P}_{\textrm{Q}}(\omega)e^{-i\omega\tau}$,
is given by
$\hat{P}_{\textrm{Q}}(\omega)=\sum_{n,m,\sigma}{\langle}\phi^\sigma_n|\hat{\sigma}_x|\phi^{\bar{\sigma}}_m{\rangle}
|\phi^\sigma_n{\rangle}{\langle}\phi^{\bar{\sigma}}_m|\delta(\omega-E^{n,\sigma}_{m,\bar{\sigma}})$,
with the energy gap $E^{n,\sigma}_{m,\bar{\sigma}}=E_{n,\sigma}-E_{m,\bar{\sigma}}$.

It should be noted that after a long-time evolution, the off-diagonal density matrix elements of the coupled qubit-resonator system are confirmed to be negligible.
{Then, the populations are decoupled from the off-diagonal elements, which restricts
the pair of projectors $\hat{P}_\mu(\omega)$ and $\hat{P}_\mu(\omega^\prime)$ in Eq.~(\ref{qme0})
to
$\hat{P}_\mu(\omega=E^\sigma_n-E^{\sigma^\prime}_{n^\prime})={\langle}\psi^\sigma_n|\hat{A}_\mu|\psi^{\sigma^\prime}_{n^\prime}{\rangle}
|\psi^\sigma_n{\rangle}{\langle}\psi^{\sigma^\prime}_{n^\prime}|$
and
$\hat{P}_\mu(\omega^\prime=E^{\sigma^\prime}_{n^\prime}-E^\sigma_n)=
{\langle}\psi^{\sigma^\prime}_{n^\prime}|\hat{A}_\mu|\psi^\sigma_n{\rangle}
|\psi^{\sigma^\prime}_{n^\prime}{\rangle}{\langle}\psi^\sigma_n|$.
Consequently, quantum master equation can be simplified to the DME
}~\cite{asettineri2018pra}
\begin{eqnarray}~\label{dme1}
\frac{d}{dt}\hat{\rho}_\textrm{S}&=&-i[\hat{H}_\textrm{S},\hat{\rho}_\textrm{S}]\nonumber\\
&&+\sum_{m,\sigma}\{\Gamma^{\textrm{R},+}_{m,\sigma}\mathcal{\hat{D}}[|\psi^\sigma_m{\rangle}{\langle}\psi^\sigma_{m-1}|]\hat{\rho}_\textrm{S}\nonumber\\
&&+\Gamma^{\textrm{R},-}_{m,\sigma}\mathcal{\hat{D}}[|\psi^\sigma_{m-1}{\rangle}{\langle}\psi^\sigma_{m}|]\hat{\rho}_\textrm{S}\}\nonumber\\
&&+\sum_{m,m^\prime,\sigma}\{\Gamma^{\textrm{Q},+}_{m,m^\prime,\sigma}
\mathcal{\hat{D}}[|\psi^\sigma_{m}{\rangle}{\langle}\psi^{\overline{\sigma}}_{m^\prime}|]\hat{\rho}_\textrm{S}\nonumber\\
&&+\Gamma^{\textrm{Q},-}_{m,m^\prime,\sigma}
\mathcal{\hat{D}}[|\psi^{\overline{\sigma}}_{m^\prime}{\rangle}{\langle}\psi^\sigma_{m}|]\hat{\rho}_\textrm{S}\},
\end{eqnarray}
{where the detailed procedures are exhibited in Appendix.}
The dressed state dissipator is given by
$\mathcal{\hat{D}}[|\psi^\sigma_m{\rangle}{\langle}\psi^{\sigma^\prime}_{l}|]\hat{\rho}_\textrm{S}
=|\psi^\sigma_m{\rangle}{\langle}\psi^{\sigma^\prime}_{l}|\hat{\rho}_\textrm{S}|\psi^{\sigma^\prime}_{l}{\rangle}{\langle}\psi^\sigma_{m}|
-\frac{1}{2}(|\psi^{\sigma^\prime}_{l}{\rangle}{\langle}\psi^{\sigma^\prime}_{l}|\hat{\rho}_\textrm{S}
+\hat{\rho}_\textrm{S}|\psi^{\sigma^\prime}_{l}{\rangle}{\langle}\psi^{\sigma^\prime}_{l}|)$, where the populations are naturally decoupled from the off-diagonal elements of the density operator.
The transition rates induced by the photon-bath interaction in Eq.~(\ref{vr}) are
\begin{subequations}
\begin{align}
\Gamma^{\textrm{R},+}_{m,\sigma}=&m(f_\sigma-g_\sigma)^2\gamma_\textrm{R}(\eta_\sigma\omega_\sigma)n_\textrm{R}(\eta_\sigma\omega_\sigma),~\label{gap}\\
\Gamma^{\textrm{R},-}_{m,\sigma}=&m(f_\sigma-g_\sigma)^2\gamma_\textrm{R}(\eta_\sigma\omega_\sigma)[1+n_\textrm{R}(\eta_\sigma\omega_\sigma)]~\label{gam}.
\end{align}
\end{subequations}
Physically, $\Gamma^{\textrm{R},+}_{m,\sigma}~(\Gamma^{\textrm{R},-}_{m,\sigma})$ describes the excitation~(relaxation) of the hybrid quantum system
from the eigenstate $|\psi^\sigma_{m-1(m)}{\rangle}$ to $|\psi^\sigma_{m(m-1)}{\rangle}$
by absorbing (releasing) the energy $\eta_\sigma\omega_\sigma$ from(to) the $\textrm{R}$th bath.
While the transition rates from the qubit-bath interaction in Eq.~(\ref{vq}) are
\begin{subequations}
\begin{align}
\Gamma^{\textrm{Q},+}_{m,m^\prime,\sigma}=&\theta(E^{m,\sigma}_{m^\prime,\overline{\sigma}})G^{m^\prime,\overline{\sigma}}_{m,\sigma}G^{m,\sigma}_{m^\prime,\overline{\sigma}}
\gamma_\textrm{Q}(E^{m,\sigma}_{m^\prime,\overline{\sigma}})n_\textrm{Q}(E^{m,\sigma}_{m^\prime,\overline{\sigma}})~\label{gqp},\\
\Gamma^{\textrm{Q},-}_{m,m^\prime,\sigma}=&\theta(E^{m,\sigma}_{m^\prime,\overline{\sigma}})G^{m^\prime,\overline{\sigma}}_{m,\sigma}G^{m,\sigma}_{m^\prime,\overline{\sigma}}
\gamma_\textrm{Q}(E^{m,\sigma}_{m^\prime,\overline{\sigma}})[1+n_\textrm{Q}(E^{m,\sigma}_{m^\prime,\overline{\sigma}})]~\label{gqm},
\end{align}
\end{subequations}
where the positive energy gap is $E^{m,\sigma}_{m^\prime,\overline{\sigma}}=E_{m,{\sigma}}-E_{m^\prime,{\overline{\sigma}}}$,
and the squeezing state overlap coefficient is defined as
$G^{n,\sigma}_{m,\bar{\sigma}}\equiv{\langle}\psi^\sigma_n|\hat{\sigma}_x|\psi^{\bar{\sigma}}_{m}{\rangle}
=\;_{a}\!{\langle}n|\exp\{-(1/2l){(\alpha_\sigma-\alpha_{\overline{\sigma}})}(\hat{a}^2-\hat{a}^{\dag^2})\}|m{\rangle}_a$,
which is specified as~\cite{pkral1990jmo}
\begin{eqnarray}~\label{gmm}
G^{m,\sigma}_{m^\prime,\bar{\sigma}}&=&\frac{({v_\sigma}/{2u_\sigma})^{m/2}}{(m!m^\prime!u_\sigma)^{1/2}}
\sum^{\min\{m,m^\prime\}}_{l=0}\frac{m^\prime!H_{m^\prime-l}}{l!(m^\prime-l)!}\frac{m!H_{m-l}}{(m-l)!}\nonumber\\
&&{\times}\left(\frac{2}{u_{\sigma}v_\sigma}\right)^{l/2}
\left(\frac{-v^*_\sigma}{2u_\sigma}\right)^{(m^\prime-l)/2},
\end{eqnarray}
with the coefficients $H_m=(-1)^{m/2}m!/(m/2)!$ for an even $m$
and $H_m=0$ for an odd $m$,
$u_\sigma=\cosh(\alpha_\sigma-\alpha_{\overline{\sigma}})$, and $v_\sigma=-\sinh(\alpha_\sigma-\alpha_{\overline{\sigma}})$.
The rate $\Gamma^{\textrm{Q},+}_{m,m^\prime,\sigma}~(\Gamma^{\textrm{Q},-}_{m,m^\prime,\sigma})$ describes the process that
the hybrid quantum system is assisted
to transit $|\psi^{\overline{\sigma}}_{m^\prime}{\rangle}~(|\psi^{{\sigma}}_{m}{\rangle})$ up (down) to
$|\psi^{{\sigma}}_{m}{\rangle}~(|\psi^{\overline{\sigma}}_{m^\prime}{\rangle})$ by exchanging energy
$E^{m,\sigma}_{m^\prime,\overline{\sigma}}$ with the $\textrm{Q}$th bath.
It need note that due to the specific selection rule of the transition factor $G^{m,\sigma}_{m^\prime,\bar{\sigma}}$ in Eq.~(\ref{gmm}),
$G^{m,\sigma}_{m^\prime,\bar{\sigma}}{\neq}0$ tightly relies on the condition $|m-m^\prime|$ to be even.
Hence, the exchange between the qubit and the $\textrm{Q}$th bath,
quantified by $\Gamma^{\textrm{Q},\pm}_{m,m^\prime,\sigma}$, occurs only between the states
$|\psi^\sigma_m{\rangle}$ and $|\psi^\sigma_{m{\pm}2l}{\rangle}$.

\subsection{Quantum master equation with FCS}
We apply the FCS to individually count the energy current into the $u$th bosonic thermal bath with the counting parameter $\chi_\mu$ after a long-time evolution,
which is based on the Markovian dressed master equation~(\ref{dme1}).
The technique of FCS was initially proposed by Levitov \emph{et al.}~\cite{levitov1992jetp,levitov1996jmp} to investigate the electron flow and fluctuations.
Technically, based on the two-time nondemolition measurement scheme, the generating function of the energy transfer is obtained as~\cite{mesp2009rmp}
\begin{eqnarray}~\label{zxt}
\mathcal{Z}(\mathrm{X},t)&{\equiv}&\textrm{Tr}\{e^{i\sum_\mu\chi_\mu\hat{H}^\mu_{\textrm{B}}(0)}e^{-i\sum_\mu\chi_\mu\hat{H}^\mu_{\textrm{B}}(t)}
\hat{\rho}_{\textrm{S}}(0)\}\nonumber\\
&=&\textrm{Tr}\{\hat{\rho}_{\mathrm{X}}(t)\},
\end{eqnarray}
where the counting parameter set is $\mathrm{X}=\{\chi_\mu\}$,
the modified density operator $\hat{\rho}_{\mathrm{X}}(t)=\hat{U}_{-\mathrm{X}}(t)\hat{\rho}_{\textrm{S}}(0)\hat{U}^\dag_{\mathrm{X}}(t)$,
the unitary evolution operator is $\hat{U}_{\mathrm{X}}(t)=\exp(-i\hat{H}_{\mathrm{X}}t)$,
the initial density operator of the hybrid system is $\hat{\rho}_{\textrm{S}}(0)$,
and the modified Hamiltonian is
$\hat{H}_{\mathrm{X}}{\equiv}\exp(i\sum_\mu\chi_\mu\hat{H}^\mu_{\textrm{B}}/2)\hat{H}\exp(-i\sum_\mu\chi_\mu\hat{H}^\mu_{\textrm{B}}/2)$,
which is specified as
$\hat{H}_{\mathrm{X}}=\hat{H}_{\textrm{S}}+\sum_\mu[\hat{H}^\mu_\textrm{B}+\hat{V}_\mu(\chi_\mu)]$,
with
$\hat{V}_{\textrm{R}}(\chi_{\textrm{R}})
=(\hat{a}^\dag+\hat{a})\sum_{k}(g_{k,\textrm{R}}e^{i\omega_k\chi_{\textrm{R}}/2}\hat{b}^\dag_{k,\textrm{R}}
+e^{-i\omega_k\chi_{\textrm{R}}/2}g^{*}_{k,\textrm{R}}\hat{b}_{k,\textrm{R}})$
and
$\hat{V}_{\textrm{Q}}(\chi_{\textrm{Q}})
=\hat{\sigma}_x\sum_{k}(g_{k,\textrm{Q}}e^{i\omega_k\chi_{\textrm{Q}}/2}\hat{b}^\dag_{k,\textrm{Q}}
+g^{*}_{k,\textrm{Q}}e^{-i\omega_k\chi_{\textrm{Q}}/2}\hat{b}_{k,\textrm{Q}})$.
Then, employing the Born-Markov approximation, the density operator is decomposed as
$\hat{\rho}_{\mathrm{X}}(t)=\hat{\rho}^{\textrm{S}}_{\mathrm{X}}(t){\otimes}\hat{\rho}_\textrm{B}$.
Within the framework of dressed master equation, we perturb $\hat{V}_{\mu}(\chi_{\mu})$ up to the second order and obtain
\begin{eqnarray}~\label{dme1}
\frac{d}{dt}\hat{\rho}^\textrm{S}_{\textrm{X}}&=&-i[\hat{H}_\textrm{S},\hat{\rho}^\textrm{S}_{\textrm{X}}]\nonumber\\
&&+\sum_{m,\sigma}\{\Gamma^{\textrm{R},+}_{m,\sigma}\mathcal{\hat{D}}_{\chi_\textrm{R}}[|\psi^\sigma_m{\rangle}{\langle}\psi^\sigma_{m-1}|]\hat{\rho}^\textrm{S}_{\textrm{X}}\nonumber\\
&&+\Gamma^{\textrm{R},-}_{m,\sigma}\mathcal{\hat{D}}_{-\chi_\textrm{R}}[|\psi^\sigma_{m-1}{\rangle}{\langle}\psi^\sigma_{m}|]\hat{\rho}^\textrm{S}_{\textrm{X}}\}\nonumber\\
&&+\sum_{m,m^\prime,\sigma}\{\Gamma^{\textrm{Q},+}_{m,m^\prime,\sigma}
\mathcal{\hat{D}}_{\chi_\textrm{Q}}[|\psi^\sigma_{m}{\rangle}{\langle}\psi^{\overline{\sigma}}_{m^\prime}|]\hat{\rho}^\textrm{S}_{\textrm{X}}\nonumber\\
&&+\Gamma^{\textrm{Q},-}_{m,m^\prime,\sigma}
\mathcal{\hat{D}}_{-\chi_\textrm{Q}}[|\psi^{\overline{\sigma}}_{m^\prime}{\rangle}{\langle}\psi^\sigma_{m}|]\hat{\rho}^\textrm{S}_{\textrm{X}}\},
\end{eqnarray}
where the modified dressed dissipator is given by
$\mathcal{\hat{D}}_{\chi}[|\psi^\sigma_m{\rangle}{\langle}\psi^{\sigma^\prime}_{l}|]\hat{\rho}^\textrm{S}_{\textrm{X}}
=e^{-i{\chi}E^{m,\sigma}_{l,\sigma^\prime}}|\psi^\sigma_m{\rangle}{\langle}\psi^{\sigma^\prime}_{l}|\hat{\rho}^\textrm{S}_{\textrm{X}}|\psi^{\sigma^\prime}_{l}{\rangle}{\langle}\psi^\sigma_{m}|
-\frac{1}{2}(|\psi^{\sigma^\prime}_{l}{\rangle}{\langle}\psi^{\sigma^\prime}_{l}|\hat{\rho}^\textrm{S}_{\textrm{X}}
+\hat{\rho}^\textrm{S}_{\textrm{X}}|\psi^{\sigma^\prime}_{l}{\rangle}{\langle}\psi^{\sigma^\prime}_{l}|)$.
Hence, we obtain the cumulant generating function at steady state as
\begin{eqnarray}
\mathcal{G}(\textrm{X})=\lim_{t{\rightarrow}\infty}\frac{1}{t}\ln[\mathcal{Z}(\textrm{X},t)].
\end{eqnarray}
Then, the $n$th cumulant into the $u$th thermal bath is given by
$J^{(n)}_u=[{\partial}^n\mathcal{G}(\textrm{X})/{\partial}(i{\chi_u})^n]{\Big{|}}_{\textrm{X}=0}$.
In the following, we focus on the current fluctuations into the $\textrm{Q}$th thermal bath.
Accordingly, the steady-state heat current, noise power, and skewness are given by
\begin{subequations}
\begin{align}
J_{ss}=&[{\partial}\mathcal{G}(\textrm{X})/{\partial}(i{\chi_{\textrm{Q}}})]{\Big{|}}_{\textrm{X}=0}~\label{j1},\\
J^{(2)}_{ss}=&[{\partial}^2\mathcal{G}(\textrm{X})/{\partial}(i{\chi_{\textrm{Q}}})^2]{\Big{|}}_{\textrm{X}=0}~\label{j2},\\
J^{(3)}_{ss}=&[{\partial}^3\mathcal{G}(\textrm{X})/{\partial}(i{\chi_{\textrm{Q}}})^3]{\Big{|}}_{\textrm{X}=0}~\label{j3}.
\end{align}
\end{subequations}
In particular, by analyzing the dynamical transition processes in Eq.~(\ref{dme1}),
the current can also be alternatively expressed as
\begin{eqnarray}~\label{current2}
J_{ss}=\sum_{m,m^\prime,\sigma}[\Gamma^{\textrm{Q},-}_{m,m^\prime,\sigma}P_{m,{\sigma}}
-\Gamma^{\textrm{Q},+}_{m,m^\prime,\sigma}P_{m^\prime,\overline{\sigma}}],
\end{eqnarray}
where $P_{m,{\sigma}}$ is the steady-state population of the eigenstate $|\psi^\sigma_m{\rangle}$.

\begin{figure}[tbp]
\includegraphics[scale=0.35]{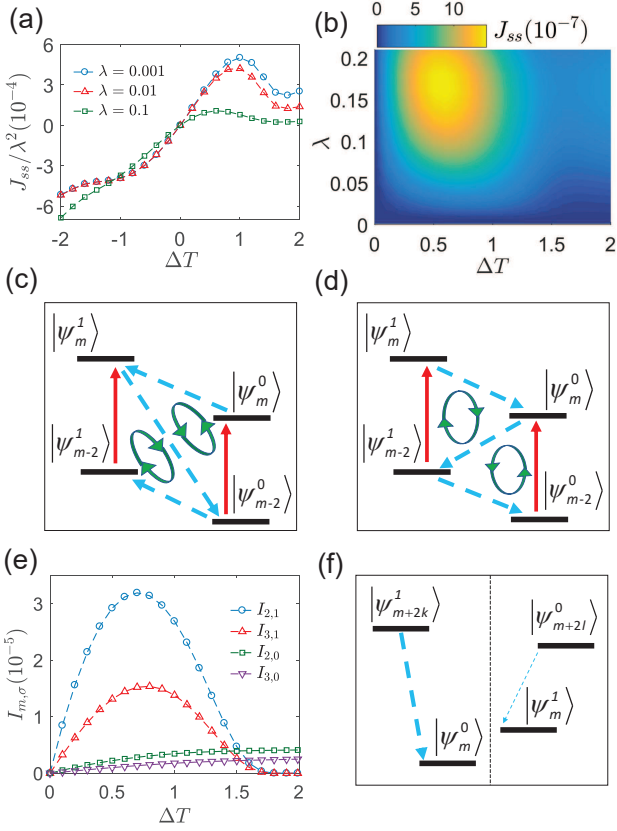}
\caption{(Color online)
(a) Influence of the temperature bias ${\Delta}T$ on the rescaled steady-state heat current $J_{ss}/\lambda^2$ from weak to strong qubit-photon hybridizations, e.g., $\lambda=0.001, 0.01$, and $0.1$;
(b) bird view of the heat current $J_{ss}$ by both modulating $\Delta{T}$ and $\lambda$;
{(c) and (d) are cyclic heat transitions to depict the first term of $I_{m,1}$ and $I_{m,0}$, respectively;}
(e) effect of the temperature bias on heat current component $I_{m,\sigma}$ given by Eqs.~(\ref{Im1}-\ref{Im0}) in the weak hybridization regime to describe the origin of the NDTC;
(f) two different types of heat transfer processes between eigenstates at strong qubit-photon hybridization,
i.e., $|\psi^1_{m+2k}{\rangle}{\rightarrow}|\psi^0_{m}{\rangle}$ and $|\psi^0_{m+2l}{\rangle}{\rightarrow}|\psi^1_{m}{\rangle}$
($m,k,l{\in}\mathrm{N}$),
where the thickness of the blue arrowed lines denoting the transition rate strengthes.
The other parameters are given by $T_{\mathrm{R}}=T_0+{\Delta}T/2$, $T_{\mathrm{Q}}=T_0-{\Delta}T/2$, $T_0=1$,
$\omega_a=1$, $\varepsilon=1$, $\alpha_{\mathrm{R}}=\alpha_{\mathrm{Q}}=10^{-3}$, and $\omega_c=10$.
}~\label{fig2}
\end{figure}


\section{Nonequilibrium thermal transport}

The NDTC is considered as one representative kind of  nonlinear effects in nonequilibrium thermal transport~\cite{bli2006apl}.
It was initially proposed by Li~\emph{et al.}~\cite{nbli2012rmp} in nonlinear phononic lattices, which is characterized as the crucial component of the thermal transistor to efficiently manage heat transport.
The NDTC describes the effect that the heat current is anomalously suppressed by increasing the bath temperature bias within two terminal setup.
In recent years, the NDTC has been extensively investigated in small quantum systems, e.g., hybrid quantum system~\cite{cwang2021cpl,jren2013prb,afornieri2016prb} and SBM~\cite{dsegal2006prb}.
However, the influence of the quadratic qubit-photon interaction on the NDTC is still illusive.

To investigate the NDTC in our system, we first show the steady-state heat current $J_{ss}/\lambda^2$~(\ref{current2}) by modulating the temperature bias
${\Delta}T$ ($T_\textrm{R}=T_0+{\Delta}T/2$ and $T_{\textrm{Q}}=T_0-{\Delta}T/2$) in Fig.~\ref{fig2}(a).
{In the negative temperature bias regime, the heat current shows trivial monotonic enhancement by increasing $|{\Delta}T|$ from weak to strong qubit-photon hybridization strengthes.
In sharp contrast, it is interesting to find that for positive ${\Delta}T$,
$J_{ss}$ always exhibits nontrivial nonmonotic behavior by enlarging ${\Delta}T$, i.e., $J_{ss}$ enhanced with small ${\Delta}T$ and suppressed at large ${\Delta}T$.
This clearly demonstrates the emergence of the NDTC.
Moreover, such novel behavior of the NDTC is bounded by $\lambda<0.25$.
Hence, we focus on the positive temperature bias in the following.
}
In Fig.~\ref{fig2}(b), we also show a comprehensive view of the cooperative effect of both the hybridization strength and the temperature bias.
The heat current forms a global peak at $\lambda{\approx}0.15$ and ${\Delta}T{\approx}0.6$.
Though not shown in the paper, it is confirmed that the NDTC persists at finite energy bias regime, i.e., $\varepsilon{\neq}\omega_a$.
It should be noted that the robustness of the NDTC in the current model over the wide qubit-photon hybridization regime is nontrivial,
which is  different from  previous results in the limiting conditions,
e.g., linearly qubit-phonon hybrid system at weak hybridization~\cite{cwang2021cpl},
nonequilibrium SBM at strong system-bath dissipation~\cite{dsegal2006prb}, and electron-magnon scattering interface with weak tunneling~\cite{jren2013prb}.
Below, we try to explore the underlying mechanism for the appearance of the NDTC in representative qubit-photon hybridization regimes.

We first analyze the expression of heat current in the weak qubit-photon hybridization case.
{
From  dynamical equation in Eq.~(\ref{dme1}) and the expression of heat current in Eq.~(\ref{current2}),
it is known that the transition rates
are fundamental to establish the dynamical energy exchange processes,
which leads to the steady-state populations and heat current.
In particular,}
the renormalization factor is reduced to
$\eta_\sigma{\approx}1-2\lambda^2/\omega^2_a$, the coefficients become
$f_\sigma{\approx}1+\lambda^2/\omega^2_a$,
$g_\sigma{\approx}(-1)^{\sigma+1}\lambda/\omega_a$,
and
$\alpha_\sigma{\approx}(-1)^{\sigma+1}\lambda/\omega_a$.
Thus, the squeezing state overlap coefficient defined in Eq.~(\ref{gmm}) is simplified as
$G^{m,\sigma}_{m^\prime,\overline{\sigma}}{\approx}\delta_{m,m^\prime}+(-1)^\sigma({\lambda}/{\omega_a})\sqrt{m^\prime(m^\prime-1)}\delta_{m,m^\prime-2}
-(-1)^\sigma({\lambda}/{\omega_a})\sqrt{m(m-1)}\delta_{m,m^\prime+2}$.
Consequently, the transition rates associated with the $\mathrm{Q}$th reservoir is given by
\begin{eqnarray}
\Gamma^{\mathrm{Q},\pm}_{m,m^\prime,\sigma}&{\approx}&\Gamma^{\mathrm{Q},\pm}_{m,m,1}\delta_{m,m^\prime}\delta_{\sigma,1}
+\Gamma^{\mathrm{Q},\pm}_{m,m-2,1}\delta_{m,m^\prime+2}\delta_{\sigma,1}\nonumber\\
&&+\Gamma^{\mathrm{Q},\pm}_{m,m+2,0}\delta_{m,m^\prime-2}\delta_{\sigma,0}.
\end{eqnarray}
{
The first rate component $\Gamma^{Q,\pm}_{m,m,1}{\approx}\gamma_Q(\pm\varepsilon)n_Q(\pm\varepsilon)$ describes the individual process
that the qubit flips up (down) assisted by the $\textrm{Q}$th bath without changing the photon state.
While the second and third components
$\Gamma^{Q,\pm}_{m,m-2,1}{\approx}m(m-1)({\lambda}/{\omega_a})^2\gamma_Q({\pm}2\omega_a\pm\varepsilon)n_Q(\pm2\omega_a\pm\varepsilon)$
and $\Gamma^{Q,\pm}_{m-2,m,0}{\approx}m(m-1)({\lambda}/{\omega_a})^2\gamma_Q(\pm2\omega_a\mp\varepsilon)n_Q(\pm2\omega_a\mp\varepsilon)$
explicitly capture the contribution of cooperative qubit-photon scattering process, i.e., the qubit flip is simultaneously accompanied by two photons emission (absorption), under the influence of both the $\textrm{R}$th and $\textrm{Q}$th thermal baths.
Hence, $\Gamma^{Q,\pm}_{m,m-2,1}$ and $\Gamma^{Q,\pm}_{m-2,m,0}$ are crucial to analytically obtain the expression of heat current.
}
Consequently, the current defined in Eq.~(\ref{current2}) with the leading order of $(\lambda/\omega_a)^2$ is expressed as
\begin{eqnarray}~\label{jss1}
J_{ss}{\approx}\frac{2\lambda^2}{\omega_a}\sum^{\infty}_{m=2}m(m-1)(I_{m,1}+I_{m,0}),
\end{eqnarray}
where the current components are
\begin{subequations}
\begin{align}
I_{m,1}=&\frac{\gamma_Q(2\omega_a+\varepsilon)e^{-\beta_\mathrm{R}m\omega_a}}
{[1+2n_\mathrm{Q}(\varepsilon)][1+n_\mathrm{R}(\omega_a)]n_\mathrm{R}(2\omega_a)}~\label{Im1}\\
&{\times}\{[1+n_\mathrm{Q}(2\omega_a+\varepsilon)]n_\mathrm{Q}(\varepsilon)n_\mathrm{R}(2\omega_a)\nonumber\\
&-n_\mathrm{Q}(2\omega_a+\varepsilon)[1+n_\mathrm{Q}(\varepsilon)][1+n_\mathrm{R}(2\omega_a)]\},\nonumber\\
I_{m,0}=&\frac{\gamma_Q(2\omega_a-\varepsilon)e^{-\beta_\mathrm{R}m\omega_a}}
{[1+2n_\mathrm{Q}(\varepsilon)][1+n_\mathrm{R}(\omega_a)]n_\mathrm{R}(2\omega_a)}~\label{Im0}\\
&{\times}\{[1+n_\mathrm{Q}(2\omega_a-\varepsilon)][1+n_\mathrm{Q}(\varepsilon)]n_\mathrm{R}(2\omega_a)\nonumber\\
&-n_\mathrm{Q}(2\omega_a-\varepsilon)n_\mathrm{Q}(\varepsilon)[1+n_\mathrm{R}(2\omega_a)]\}.\nonumber
\end{align}
\end{subequations}
{
Specifically, the component $I_{m,1}$ is contributed by two competing cyclic transitions,
where the first term characterized as $[1+n_\mathrm{Q}(2\omega_a+\varepsilon)]n_\mathrm{Q}(\varepsilon)n_\mathrm{R}(2\omega_a)$ is specified by the transition paths
\begin{eqnarray}
&&|\psi^1_m{\rangle}{\rightarrow}|\psi^0_{m-2}{\rangle}{\rightarrow}|\psi^1_{m-2}{\rangle}{\rightarrow}|\psi^1_m{\rangle},\\
&&|\psi^1_m{\rangle}{\rightarrow}|\psi^0_{m-2}{\rangle}{\rightarrow}|\psi^0_m{\rangle}{\rightarrow}|\psi^1_{m}{\rangle}.\nonumber
\end{eqnarray}
And the microscopic physical pictures are also graphed in Fig.~\ref{fig2}(c).
Accordingly, the second term can be straightforwardly obtained by reversing the transition directions of Fig.~\ref{fig2}(c).
In analogy, $I_{m,0}$ is also composed by opposite cyclic components, where the first term
proportional to $[1+n_\mathrm{Q}(2\omega_a-\varepsilon)][1+n_\mathrm{Q}(\varepsilon)]n_\mathrm{R}(2\omega_a)$
is contributed by transition paths
\begin{eqnarray}
&&|\psi^0_m{\rangle}{\rightarrow}|\psi^1_{m-2}{\rangle}{\rightarrow}|\psi^0_{m-2}{\rangle}{\rightarrow}|\psi^0_m{\rangle},\\
&&|\psi^0_m{\rangle}{\rightarrow}|\psi^1_{m-2}{\rangle}{\rightarrow}|\psi^1_m{\rangle}{\rightarrow}|\psi^0_{m}{\rangle},\nonumber
\end{eqnarray}
and physically depicted in Fig.~\ref{fig2}(d).
}

{
Then, we analyze the mechanism for emergence of the NDTC with weak qubit-resonator hybridization.
At weak temperature bias limit, the heat current Eq.~(\ref{jss1}) behaves $J_{ss}{\propto}{\Delta}T$,
which denotes that $J_{ss}$ shows linear enhancement by increasing ${\Delta}T$.
While in the large ${\Delta}T$ regime, e.g., $T_{\textrm{R}}{\approx}2T_0$ and $T_{\textrm{Q}}{\approx}0$,
the  directional transitions, i.e., $|\psi^0_{m-2}{\rangle}{\rightarrow}|\psi^1_{m-2}{\rangle}$,
$|\psi^0_{m-2}{\rangle}{\rightarrow}|\psi^0_{m}{\rangle}$ and
$|\psi^0_{m-2}{\rangle}{\rightarrow}|\psi^1_{m}{\rangle}$,
are dramatic suppressed.
Then, all cyclic transitions to compose $I_{m,1}$ break down, which naturally leads to $I_{m,1}{\approx}0$.
Hence, $I_{m,1}$ clearly exhibits the nonmonotonic behavior by increasing the bias ${\Delta}T$.
Moreover, the magnitude of $I_{m,1}$ is  larger than $I_{m,0}$ over a wide temperature bias regime ($0<{\Delta}T{<}1.3$),
with $m=2,3$ exemplified in Fig.~\ref{fig2}(e).
Therefore, we conclude that the effect of the NDTC originates from the suppression of cyclic components of $I_{m,1}$.
However, the finite residue of the heat current at large temperature bias limit results from the robustness of $I_{m,0}$,
of which all cyclic components show monotonic enhancement by increasing ${\Delta}T$.
}

{
Next, we turn to investigate the mechanism of the NDTC beyond weak qubit-photon hybridization.
Due to multi-photon transfer processes, it is rather difficult to obtain the analytical expression of heat current.
However, we may resort to transition rates $\Gamma^{\pm}_{m,m^\prime,\sigma}$ for the physical insight of energy exchange processes.
In particular, in large hybridization regime the coefficient $G^{m,\sigma}_{m^\prime,\overline{\sigma}}$ Eq.~(\ref{gmm})
will be strongly suppressed with the increase of number difference $|m-m^\prime|$.
Combining the dramatic frequency mismatch of bosonic squeezing modes Eq.~(\ref{hs1}), i.e., $\eta_1\omega_1{\gg}\eta_0\omega_0$,
the directional transition from $|\psi^0_{m+2l}{\rangle}$ and $|\psi^{1}_{m}{\rangle}$ is almost inhibited between squeezing states $|\psi^0_{m}{\rangle}$ and $|\psi^{1}_{m^\prime}{\rangle}$ [shown in Fig.~\ref{fig2}(f)],
which is restricted by $l{\gg}1$ to overcome the positive energy bias $E^{0}_{m+2l}{>}E^1_{m}$.
On the contrary,  the transition from $|\psi^1_{m+2k}{\rangle}$ and $|\psi^{0}_{m}{\rangle}$ is generally allowed
under the condition $E^{1}_{m+2k}{>}E^0_{m}$.
Hence, the transition rate $\Gamma^{\textrm{Q},\pm}_{m+2l,m,0}$ becomes much smaller than $\Gamma^{\textrm{Q},\pm}_{m+2k,m,1}$ with $l{\gg}k$ over a wide temperature bias regime,
as schemed in Fig.~\ref{fig2}(f).
Consequently, at large temperature limit (e.g., ${\Delta}T{\approx}2T_0$) the qubit is almost localized to the ground state $|0{\rangle}$.
Such localization of the qubit apparently blocks the energy exchange between the qubit and the $\textrm{Q}$th thermal bath,
which finally results in the emergence of the NDTC.
}





\begin{figure}[tbp]
\includegraphics[scale=0.55]{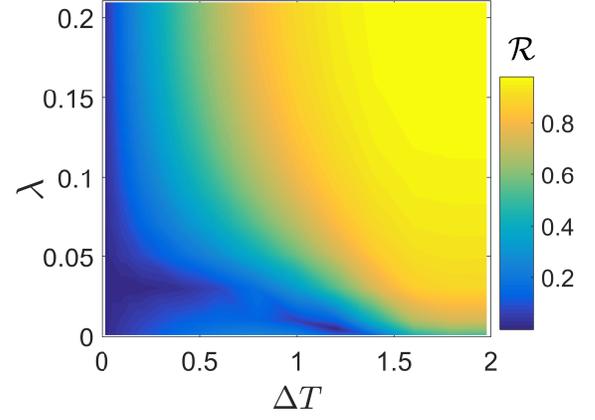}
\caption{(Color online) Thermal rectification factor $\mathcal{R}$ defined in Eq.~(\ref{tr0}) versus the temperature bias ${\Delta}T$ and the qubit-photon hybridization strength $\lambda$.
The other parameters are the same as those in Fig.~\ref{fig2}.
}~\label{fig3}
\end{figure}
The novel thermal transport effect in this system can also be characterized by the thermal rectification effect, which
in analogy with the electronic diode, was theoretically proposed by Li \emph{et al.}~\cite{bli2004prl} to study the heat flux asymmetry in nonlinear phononic lattices.
In recent years, the concept of thermal rectification has been extensively investigated in various hybrid quantum setups~\cite{cwang2021cpl,dsegal2005prl,mjmperez2015nn}.
{The thermal rectification is applied to characterize nonreciprocity of the nonequilibrium system,
that within the two terminal setup heat current is larger in one direction than the opposite counterpart by only exchanging two bath temperatures.}
Quantitatively, the thermal rectification is defined by the factor~\cite{dsegal2005prl,lfzhang2009prb}
\begin{eqnarray}~\label{tr0}
\mathcal{R}=\frac{|J_{ss}({\Delta}T)+J_{ss}(-{\Delta}T)|}{\max\{|J_{ss}({\Delta}T)|,|J_{ss}(-{\Delta}T)|\}},
\end{eqnarray}
where $J_{ss}(\pm{\Delta}T)$ denote the current under the conditions
$T_\mathrm{R}=T_0\pm\Delta{T}/2$ and $T_\mathrm{Q}=T_0\mp\Delta{T}/2$.
From the definition in Eq.~(\ref{tr0}), it is known that the rectification factor shows symmetric behavior as $\mathcal{R}(\Delta{T})=\mathcal{R}(-\Delta{T})$.
{Moreover, the thermal rectification becomes significant in the upper limit $\mathcal{R}=1$,
where the extreme nonreciprocity occurs as $|J_{ss}({\Delta}T)|{\gg}|J_{ss}(-{\Delta}T)|$.
While the rectification disappears as $\mathcal{R}=0$, which denotes the complete reciprocity behavior, i.e.,
$J_{ss}(-{\Delta}T){=}-J_{ss}({\Delta}T)$.
}

{
We plot Fig.~\ref{fig3} to investigate the behavior of thermal rectification by tuning the temperature bias ${\Delta}T$.
In small temperature bias regime, the rectification factor $\mathcal{R}$ generally keeps finite from weak to strong qubit-photon hybridizations.
It results from the intrinsic nonreciprocity of the qubit-resonator hybrid system, owning the spin and boson nature respectively.
Quantitatively, we may obtain some insight from Eq.~(\ref{jss1}) at weak qubit-resonator hybridization.
By exchanging thermal bath temperatures, the cyclic transition weight of current components are significantly changed.
For instance, the exchanged cyclic kernel in $I_{m,1}$ is given by
$\{[1+n_\mathrm{R}(2\omega_a+\varepsilon)]n_\mathrm{R}(\varepsilon)n_\mathrm{Q}(2\omega_a)
-n_\mathrm{R}(2\omega_a+\varepsilon)[1+n_\mathrm{R}(\varepsilon)][1+n_\mathrm{Q}(2\omega_a)]\}$,
which is distinct from the counterpart $\{[1+n_\mathrm{Q}(2\omega_a+\varepsilon)]n_\mathrm{Q}(\varepsilon)n_\mathrm{R}(2\omega_a)
-n_\mathrm{Q}(2\omega_a+\varepsilon)[1+n_\mathrm{Q}(\varepsilon)][1+n_\mathrm{R}(2\omega_a)]\}$ in Eq.~(\ref{Im1}).
This may partially explains the existence of finite thermal rectification.
While in large temperature bias regime, the rectification factor is further enhanced.
In particular at strong hybridization (e.g., $\lambda{=}0.1$), the factor approaches the unit ($\mathcal{R}{\rightarrow}1$), and keeps robust.
Such enhancement results from emergence of the NDTC only at positive ${\Delta}T$,
i.e., $-J_{ss}(\Delta{T}=-2){\gg}J_{ss}(\Delta{T}=2)$, as illustrated in Fig.~\ref{fig2}(a).}
Therefore, the qubit-resonator hybrid system with quadratic interaction may provide a potential platform
to realize the efficient thermal diode.


\begin{figure}[tbp]
\includegraphics[scale=0.45]{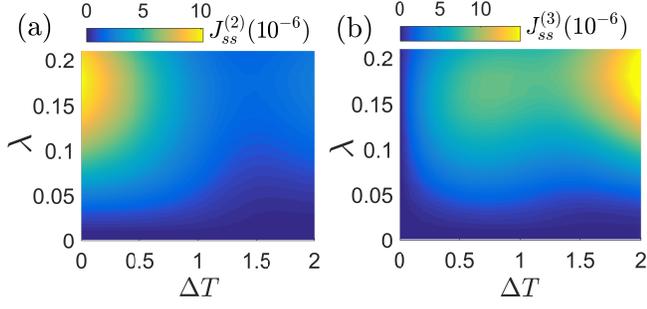}
\caption{(Color online)
(a) Noise power $J^{(2)}_{ss}$ and (b) skewness $J^{(3)}_{ss}$ defined in Eqs.~(\ref{j2}-\ref{j3}) by tuning the bath temperature bias ${\Delta}T$ and qubit-photon hybridization strength $\lambda$.
The other parameters are the same as those given in Fig.~\ref{fig2}.
}~\label{fig4}
\end{figure}

The higher-order thermal transport effect can be characterized by investigating the steady-state current fluctuations,
which are typically quantified by the noise power and skewness.
{The noise power $J^{(2)}_{ss}$ in Eq.~(\ref{j2}) characterizes zero-frequency power spectrum of the heat noise generated by the stochastic heat exchange between the hybrid system and $\mathrm{Q}$th bath,
which is generally not directly related with the heat current at small quantum systems~\cite{levitov1992jetp,mesp2009rmp,fzhan2011prb}.
Thus, the noise power is considered as one indispensable factor to quantify the current fluctuation.}
It is interesting to find that $J^{(2)}_{ss}$ shows global peak with small temperature bias at strong hybridization ($\lambda{\approx}0.18$),
whereas it is generally suppressed as the temperature bias increases, as shown in Fig.~\ref{fig4}(a).
Hence, the monotonic suppression behavior of the noise power by increasing ${\Delta}T$ is distinct from the steady-state heat current.
{While the finite skewness $J^{(3)}_{ss}$ defined in Eq.~(\ref{j3}) characterizes  asymmetric probability distribution of the transferred heat energy around the mean value,
whereas the vanishing skewness corresponds to a symmetric Gaussian distribution~\cite{rbelousov2016pre}.
The dramatic positive skewness can be found at strong qubit-photon hybridization, which exhibits that the probability distribution of the transferred heat
deviates from the Gaussian case.
Furthermore, it is significantly enhanced at large bias limit (${\Delta}T=2$) , as shown in Fig.~\ref{fig4}(b).
}
Therefore, the current fluctuations exhibit nontrivial features compared to the heat current, which fertilize
heat transport pictures of the hybrid quantum system.

{
Via the Fourier transformation
the generating function Eq.~(\ref{zxt}) to count heat flow into the $\mu$th thermal bath, can be re-expressed as~\cite{nasinitsyn2007epl,jren2010prl}
$\mathcal{Z}_t(\chi_\mu)=\int^\infty_{-\infty}d{Q}P_t(Q)e^{iQ\chi_\mu}$,
where $P_t(Q)$ denotes the conditional probability to transport heat energy $Q$ into the $\mu$th bath by the time $t$.
Then, the cumulant generating function at steady state is obtained as  $\mathcal{G}(\chi_\mu)=\lim_{t{\rightarrow}\infty}\frac{1}{t}\ln[\mathcal{Z}_t(\chi_\mu)]$.
Consequently, the noise power and skewness are given by
$J^{(2)}_{ss}=\lim_{t{\rightarrow}\infty}\frac{1}{t}[{\langle}(Q-{\langle}Q{\rangle})^2{\rangle}]$
and
$J^{(3)}_{ss}=\lim_{t{\rightarrow}\infty}\frac{1}{t}[{\langle}(Q-{\langle}Q{\rangle})^3{\rangle}]$,
where ${\langle}Q^n{\rangle}=\int^\infty_{-\infty}d{Q}Q^nP_t(Q)/\mathcal{Z}_t(0)$.
From the practical view, $Q=\int^t_0{J(\tau)d\tau}$ could be experimentally detected via the trajectory measurement~\cite{yang2020nc},
with $J(\tau)$ the instantaneous heat current.
By collecting many trajectories of quantum heat transport after long-time evolution,
the distribution $P_{t{\rightarrow}\infty}(Q)$ can be approximately obtained.
Under such measurement scheme, noise power and skewness could be experimentally measured.
}

\section{Stationary Photon quadrature squeezing}

\begin{figure}[tbp]
\includegraphics[scale=0.5]{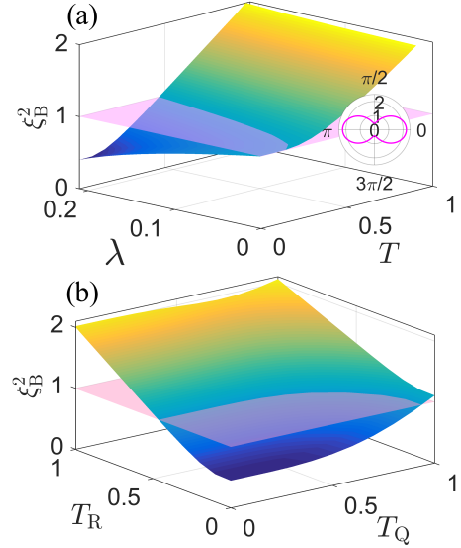}
\caption{(Color online)
Photon quadrature squeezing factor $\xi^2_\textrm{B}$ defined in Eq.~(\ref{xi2})
(a) at equilibrium $T_\mathrm{R}=T_\mathrm{Q}=T$, with the inset showing the angle distribution of $({\Delta}\hat{X}_\theta)^2$ at $\lambda=0.2$,
and (b) far-from-equilibrium by tuning two bath temperatures $T_\mathrm{R}$ and $T_\mathrm{Q}$ individually, with $\lambda=0.2$.
The other parameters are the same as those given in Fig.~\ref{fig2}.
}~\label{fig5}
\end{figure}
Quadrature squeezing is considered as one important nonclassical effect,
where the noise of  a single bosonic quadrature breaks the Heisenberg fluctuation limit at vacuum~\cite{dstoler1970prd,dstoler1971prd,ula2016ps}.
Typically, from the dynamical view the noise is suppressed at some momentum accompanied by the enhancement of noise at other times~\cite{jrk1988pra,ax2020np}.
While from the stationary view (i.e., ground state or steady state), the noise of one bosonic quadrature component is below the zero-point level,
whereas the other quadrature one surpasses the value in the vacuum state~\cite{as2013prl,ak2013pra,eew2015science}.
Generally, the quadrature squeezing factor can be defined as~\cite{jma2011pr}
{
\begin{eqnarray}~\label{xi2}
\xi^2_\textrm{B}=\min_{\theta{\in}[0,2\pi)}({\Delta}\hat{X}_\theta)^2
\end{eqnarray}
}
to quantify the photon squeezing at steady state,
where $\hat{X}_\theta=\hat{X}\cos\theta+\hat{P}\sin\theta$,
with the canonical position operator $\hat{X}=\hat{a}^\dag+\hat{a}$
and the canonical momentum operator $\hat{P}=i(\hat{a}^\dag-\hat{a})$.
If $\xi^2_\textrm{B}<1$, the photon is squeezed,
whereas the squeezing effect vanishes as $\xi^2_\textrm{B}>1$.
While for the photon mode at the coherent state, the factor
becomes the unit.
The concept of quadrature squeezing has been extensively applied
in various hybrid quantum setups, e.g., optomechanics~\cite{ak2013pra,eew2015science,jqliao2011pra,xylu2015pra} and quantum electrodynamics systems (QEDs)~\cite{mm2008prl,jkxie2020pra}.
Here, we investigate the steady-state photon quadrature squeezing under the influence of two bath temperatures and the qubit-photon hybridization strength.


We first study the photon squeezing at equilibrium in Fig.~\ref{fig5}(a), i.e., $T_{\mathrm{R}}=T_{\mathrm{Q}}=T$.
It is found that the squeezing effect ($\xi^2_{\textrm{B}}<1$) occurs in the low-temperature regime,
and is gradually enhanced as the qubit-photon hybridization strength increases.
By analyzing the angle distribution of $\xi^2_{\textrm{B}}<1$ in the squeezing regime [e.g., $\lambda=0.2$ and $T=0$ in the inset of Fig.~\ref{fig5}(a)],
we find that the noise of one quadrature component $\hat{P}$ is significantly reduced,
whereas the other component $\hat{X}$ fails to squeeze.
Moreover, we investigate the effect of bath temperatures bias on the squeezing factor at strong hybridization ($\lambda=0.2$) in Fig.~\ref{fig5}(b).
The squeezing factor is clearly shown asymmetrically driven by two bath temperatures.
Particularly, in the low temperature of $T_\mathrm{R}$, e.g., $T_\mathrm{R}=0$, the squeezing effect persists as $T_\mathrm{Q}<0.9$.
Therefore, we conclude that the photon squeezing effect favors strong qubit-photon hybridization,
and persists even at comparatively high temperature of $T_{\mathrm{Q}}$.

\section{Conclusion}

In conclusion, we studied the steady-state heat flow and photon quadrature squeezing in an open qubit-resonator quantum system,
where the optical resonator is quadratically coupled to the qubit, with each individually interacting with the bosonic thermal bath.
We applied the dressed master equation combined with the full counting statistics to quantify the heat current and current fluctuations,
which is applicable after long time evolution with negligible off-diagonal elements of the system density matrix in the squeezed states basis
$\{|\psi^\sigma_m{\rangle}\}$.
The transition rates $\Gamma^{\mathrm{R},\pm}_{m,\sigma}$ and $\Gamma^{\mathrm{Q},\pm}_{m,m^\prime,\sigma}$, characterizing the energy exchange between the hybrid quantum system components (optical resonator and qubit)
and the corresponding thermal baths ($\mathrm{R}$th and $\mathrm{Q}$th), are expressed at Eqs.~(\ref{gap}-\ref{gam}) and Eqs.~(\ref{gqp}-\ref{gqm}),
respectively.
In particular, the rates $\Gamma^{\mathrm{Q},\pm}_{m,m^\prime,\sigma}$ are tightly related with
the squeezed state overlap coefficient $G^{m,\sigma}_{m^\prime,\overline{\sigma}}$ in Eq.~(\ref{gmm}),
which clarifies the specific transition rule between squeezed states with opposite spin branches.

We  investigated the steady-state heat current by both tuning the bath temperature bias and the qubit-resonator hybridization strength.
It is interesting to find that the heat current always shows nonmonotonic behavior by increasing the temperature bias,
i.e., the NDTC, from weak to strong hybridizations.
We also exploited the underlying mechanism to exhibit the NDTC.
At weak hybridization, the heat current is analytically contributed by two components $I_{m,\sigma}$ in Eqs.~(\ref{Im1}-\ref{Im0}),
which are both described by the cyclic heat exchange processes.
And the suppression of cyclic heat transitions of $I_{m,1}$ dominates the emergence of the NDTC.
While at strong hybridization, the almost directional transitions from the squeezed states branch $\{|\psi^1_m{\rangle}\}$ to $\{|\psi^0_m{\rangle}\}$
mainly contribute to the heat transfer at finite temperature bias, due to the large frequency mismatch of two squeezed photon modes in Eq.~(\ref{hs1}).
Then, we investigated the thermal rectification effect in Eq.~(\ref{tr0}),
which becomes highly efficient at strong qubit-resonator hybridization and large temperature bias.
The intrinsic nonreciprocity of the qubit-resonator hybrid system and the emergence of the NDTC contribute to
such  highly thermal rectification.
Moreover, we analyzed steady-state noise power and skewness, which show global peaks with large hybridization strength
at small and large temperature bias limits, respectively.
These novel fluctuation features fertilize heat transport properties of the hybrid system.

We also investigated the photon quadrature squeezing effect quantified by the generalized factor $\xi^2_{\mathrm{B}}$ defined in Eq.~(\ref{xi2}).
At equilibrium ($T_{\mathrm{R}}=T_{\mathrm{Q}}=T$), the squeezing behavior occurs in the comparative low-temperature regime,
and becomes enhanced with the increase of the qubit-resonator hybridization strength.
The variance of the momentum operator $\hat{P}$, corresponding to $\theta=\pi/2$, is dramatically suppressed,
whereas the other quadrature component $\hat{X}$($\theta=0$) becomes noisy.
While out of equilibrium ($T_{\mathrm{R}}{\neq}T_{\mathrm{Q}}$) and strong hybridization,
the squeezing factor shows asymmetric behavior by tuning two bath temperatures.

We hope that the quadratically coupled qubit-resonator hybrid quantum system can be considered as one potential hybrid platform
to realize efficient quantum thermal devices, e.g., the NDTC and thermal rectifier, and nonclassical behavior of  photons, e.g., quadrature squeezing.

\section*{ACKNOWLEDGEMENTS}

C.W. is supported by the
National Natural Science Foundation
of China under Grant No. 11704093
and the Opening Project of Shanghai Key Laboratory of Special Artificial Microstructure Materials and Technology.
H.C. acknowledges the National Natural Science Foundation
of China under Grant No. 11704338.
J.-Q.L. is supported in part by National Natural Science Foundation of
China (Grants No. 11822501, No. 11774087, and No.
11935006), Hunan Science and Technology Plan Project
(Grant No. 2017XK2018), and the Science and Technology Innovation Program of Hunan Province (Grant No.
2020RC4047).

{
\section*{Appendix: Derivation of the quantum dressed master equation}
Based on the Born approximation, the density operator of the total system is decomposed by
$\hat{\rho}_{\textrm{tot}}{\approx}\hat{\rho}_\textrm{S}{\otimes}\hat{\rho}_\textrm{B}$,
where $\hat{\rho}_\textrm{S}$ is the density operator of the qubit-resonator  hybrid system,
and $\hat{\rho}_\textrm{B}=\Pi_{u=\textrm{R},\textrm{Q}}\exp[-\hat{H}^u_\textrm{B}/(k_BT_u)]/Z$ is the density operator of thermal baths at equilibrium, with $Z=\textrm{Tr}_\textrm{B}\{\Pi_{u=\textrm{R},\textrm{Q}}\exp[-\hat{H}^u_\textrm{B}/(k_BT_u)]\}$ the partition function.
Then, we perturb system-bath interactions, i.e., $\hat{V}_{\textrm{R}}$ and $\hat{V}_{\textrm{Q}}$, to obtain the quantum  master equation as
\begin{eqnarray}
\frac{d}{dt}\hat{\rho}_{\textrm{S}}&=&i[\hat{\rho}_{\textrm{S}},\hat{H}_{\textrm{S}}]\\
&&+\sum_{\mu=\textrm{R},\textrm{Q}} \int^\infty_0{d\tau} \textrm{Tr}_{\textrm{B}}\{[\hat{V}_\mu,[\hat{V}_\mu(-\tau),\hat{\rho}_{\textrm{S}}{\otimes}\hat{\rho}_{\textrm{B}}]]\},\nonumber
\end{eqnarray}
where the commutation relation is $[\hat{X},\hat{Y}]=\hat{X}\hat{Y}-\hat{Y}\hat{X}$,
and $\textrm{Tr}_{\textrm{B}}$ denotes the trace over the degrees of freedom of thermal baths.
Moreover, since the interaction term is expressed as $\hat{V}_{\mu}=\hat{A}_\mu{\otimes}\hat{B}_\mu$,
with $\hat{A}_{\textrm{R}}=(\hat{a}^\dag+\hat{a})$, $\hat{A}_{\textrm{Q}}=\hat{\sigma}_x$,
and $\hat{B}_{\mu}=\sum_k(g_{k,\mu}\hat{b}^\dag_{k,\mu}+g^{*}_{k,\mu}\hat{b}_{k,\mu})~(\mu=\textrm{R},\textrm{S})$,
quantum master equation is specified as
\begin{eqnarray}
\frac{d}{dt}\hat{\rho}_{\textrm{S}}&=&i[\hat{\rho}_{\textrm{S}},\hat{H}_{\textrm{S}}]
+\sum_{\mu=\textrm{R},\textrm{Q}}\int^\infty_0{d\tau}{\times} \\
&&\{C_\mu(\tau)[\hat{A}_\mu(-\tau)\hat{\rho}_{\textrm{S}},\hat{A}_\mu]+\mathrm{h.c.}\},\nonumber
\end{eqnarray}
where the correlation function is
$C_\mu(\tau)\equiv\langle\hat{B}_\mu(\tau)\hat{B}_\mu(0)\rangle =\sum_k |g_{k,\mu}|^2[n_{k,\mu}e^{i\omega_{k,\mu}\tau}+(1+n_{k,\mu})e^{-i\omega_{k,\mu}\tau}]$,
and the Bose-Einstein distribution function is $n_{k,\mu}=1/[\exp(\omega_{k,\mu}/k_BT_\mu)-1]$.
Then, under the eigenbasis of $\hat{H}_{\textrm{S}}$,
i.e. $\hat{H}_{\textrm{S}}|\psi^\sigma_n{\rangle}=E^\sigma_n|\psi^\sigma_n{\rangle}$,
we expand the operator
\begin{eqnarray}
\hat{A}_\mu(\tau)=\sum_{n,n^\prime,\sigma,\sigma^\prime}e^{i(E^\sigma_n-E^{\sigma^\prime}_{n^\prime})\tau}
{\langle}\psi^\sigma_n|\hat{A}_\mu|\psi^{\sigma^\prime}_{n^\prime}{\rangle}
|\psi^\sigma_n{\rangle}{\langle}\psi^{\sigma^\prime}_{n^\prime}|.
\end{eqnarray}
If we reexpress
$\hat{A}_\mu(\tau)=\sum_{\omega=E^\sigma_n-E^{\sigma^\prime}_{n^\prime}}\hat{P}_\mu(\omega)e^{i\omega\tau}$,
the quantum master equation can be obtained as
\begin{eqnarray}~\label{appendix-qme0}
\frac{d}{dt}\hat{\rho}_\textrm{S}&=&i[\hat{\rho}_\textrm{S},\hat{H}_\textrm{S}]
+\frac{1}{2}\sum_{\omega,\omega^\prime;u=\textrm{R},\textrm{Q}}
\{\kappa_u(\omega^\prime)[\hat{P}_u(\omega^\prime)\hat{\rho}_s,\hat{P}_u(\omega)]\nonumber\\
&&+\mathrm{h.c.}\},
\end{eqnarray}
where the rates are $\kappa_u(\omega^\prime)=\gamma_u(\omega^\prime)n_u(\omega^\prime)$.
The Eq.~(\ref{appendix-qme0}) is identical with Eq.~(\ref{qme0}).

From Eq.~(\ref{appendix-qme0}), it is known that the populations are generally coupled with the off-diagonal elements of $\hat{\rho}_\textrm{S}$ at finite-time dynamics.
However, after long-time evolution the off-diagonal elements become negligible due to the full thermalization of the qubit-resonator hybrid system.
Hence, the populations dynamics are approximately decoupled from the off-diagonal elements.
Specifically, the projectors $\hat{P}_\mu(\omega)$ and $\hat{P}_\mu(\omega^\prime)$ in Eq.~(\ref{appendix-qme0})
are restricted to
$\hat{P}_\mu(\omega=E^\sigma_n-E^{\sigma^\prime}_{n^\prime})={\langle}\psi^\sigma_n|\hat{A}_\mu|\psi^{\sigma^\prime}_{n^\prime}{\rangle}
|\psi^\sigma_n{\rangle}{\langle}\psi^{\sigma^\prime}_{n^\prime}|$
and
$\hat{P}_\mu(\omega^\prime=E^{\sigma^\prime}_{n^\prime}-E^\sigma_n)=
{\langle}\psi^{\sigma^\prime}_{n^\prime}|\hat{A}_\mu|\psi^\sigma_n{\rangle}
|\psi^{\sigma^\prime}_{n^\prime}{\rangle}{\langle}\psi^\sigma_n|$.
Consequently, the quantum dressed master equation can be obtained as
\begin{eqnarray}~\label{appendix-dme1}
\frac{d}{dt}\hat{\rho}_\textrm{S}&=&-i[\hat{H}_\textrm{S},\hat{\rho}_\textrm{S}]\nonumber\\
&&+\sum_{m,\sigma}\{\Gamma^{\textrm{R},+}_{m,\sigma}\mathcal{\hat{D}}[|\psi^\sigma_m{\rangle}{\langle}\psi^\sigma_{m-1}|]\hat{\rho}_\textrm{S}\nonumber\\
&&+\Gamma^{\textrm{R},-}_{m,\sigma}\mathcal{\hat{D}}[|\psi^\sigma_{m-1}{\rangle}{\langle}\psi^\sigma_{m}|]\hat{\rho}_\textrm{S}\}\nonumber\\
&&+\sum_{m,m^\prime,\sigma}\{\Gamma^{\textrm{Q},+}_{m,m^\prime,\sigma}
\mathcal{\hat{D}}[|\psi^\sigma_{m}{\rangle}{\langle}\psi^{\overline{\sigma}}_{m^\prime}|]\hat{\rho}_\textrm{S}\nonumber\\
&&+\Gamma^{\textrm{Q},-}_{m,m^\prime,\sigma}
\mathcal{\hat{D}}[|\psi^{\overline{\sigma}}_{m^\prime}{\rangle}{\langle}\psi^\sigma_{m}|]\hat{\rho}_\textrm{S}\},
\end{eqnarray}
which is the same as Eq.~(\ref{dme1}).
The dressed state dissipator is given by
$\mathcal{\hat{D}}[|\psi^\sigma_m{\rangle}{\langle}\psi^{\sigma^\prime}_{l}|]\hat{\rho}_\textrm{S}
=|\psi^\sigma_m{\rangle}{\langle}\psi^{\sigma^\prime}_{l}|\hat{\rho}_\textrm{S}|\psi^{\sigma^\prime}_{l}{\rangle}{\langle}\psi^\sigma_{m}|
-\frac{1}{2}(|\psi^{\sigma^\prime}_{l}{\rangle}{\langle}\psi^{\sigma^\prime}_{l}|\hat{\rho}_\textrm{S}
+\hat{\rho}_\textrm{S}|\psi^{\sigma^\prime}_{l}{\rangle}{\langle}\psi^{\sigma^\prime}_{l}|)$.
The transition rates involved with the resonator-bath interaction are described as
\begin{subequations}
\begin{align}
\Gamma^{\textrm{R},+}_{m,\sigma}=&m(f_\sigma-g_\sigma)^2\gamma_\textrm{R}(\eta_\sigma\omega_\sigma)n_\textrm{R}(\eta_\sigma\omega_\sigma),\\
\Gamma^{\textrm{R},-}_{m,\sigma}=&m(f_\sigma-g_\sigma)^2\gamma_\textrm{R}(\eta_\sigma\omega_\sigma)[1+n_\textrm{R}(\eta_\sigma\omega_\sigma)].
\end{align}
\end{subequations}
And the transition rates related with the qubit-bath interaction  are given by
\begin{subequations}
\begin{align}
\Gamma^{\textrm{Q},+}_{m,m^\prime,\sigma}=&\theta(E^{m,\sigma}_{m^\prime,\overline{\sigma}})G^{m^\prime,\overline{\sigma}}_{m,\sigma}G^{m,\sigma}_{m^\prime,\overline{\sigma}}
\gamma_\textrm{Q}(E^{m,\sigma}_{m^\prime,\overline{\sigma}})n_\textrm{Q}(E^{m,\sigma}_{m^\prime,\overline{\sigma}}),\\
\Gamma^{\textrm{Q},-}_{m,m^\prime,\sigma}=&\theta(E^{m,\sigma}_{m^\prime,\overline{\sigma}})G^{m^\prime,\overline{\sigma}}_{m,\sigma}G^{m,\sigma}_{m^\prime,\overline{\sigma}}
\gamma_\textrm{Q}(E^{m,\sigma}_{m^\prime,\overline{\sigma}})[1+n_\textrm{Q}(E^{m,\sigma}_{m^\prime,\overline{\sigma}})],
\end{align}
\end{subequations}
where the positive energy gap is $E^{m,\sigma}_{m^\prime,\overline{\sigma}}=E_{m,{\sigma}}-E_{m^\prime,{\overline{\sigma}}}$,
and the squeezing state overlap coefficient is defined as
$G^{n,\sigma}_{m,\bar{\sigma}}\equiv{\langle}\psi^\sigma_n|\hat{\sigma}_x|\psi^{\bar{\sigma}}_{m}{\rangle}$,
which is specified as
\begin{eqnarray}
G^{m,\sigma}_{m^\prime,\bar{\sigma}}&=&\frac{({v_\sigma}/{2u_\sigma})^{m/2}}{(m!m^\prime!u_\sigma)^{1/2}}
\sum^{\min\{m,m^\prime\}}_{l=0}\frac{m^\prime!H_{m^\prime-l}}{l!(m^\prime-l)!}\frac{m!H_{m-l}}{(m-l)!}\nonumber\\
&&{\times}\left(\frac{2}{u_{\sigma}v_\sigma}\right)^{l/2}
\left(\frac{-v^*_\sigma}{2u_\sigma}\right)^{(m^\prime-l)/2},
\end{eqnarray}
with the coefficients $H_m=(-1)^{m/2}m!/(m/2)!$ for an even $m$
and $H_m=0$ for an odd $m$,
$u_\sigma=\cosh(\alpha_\sigma-\alpha_{\overline{\sigma}})$, and $v_\sigma=-\sinh(\alpha_\sigma-\alpha_{\overline{\sigma}})$.
}


\begin{thebibliography}{99}
\bibitem{gchen2005book}G. Chen, \emph{Nanoscale energy transport and conversion} (Oxford University Press, England, 2005).

\bibitem{gbenenti2017pr}G. Benenti, G. Casati, K. Saito, and R. S. Whitney, Phys. Rep. \textbf{694}, 1 (2017).
\bibitem{mesposito2019prl}K. Ptaszy\'{n}ski and M. Esposito, Phys. Rev. Lett. \textbf{122}, 150603 (2019).

\bibitem{aronzani2018np}A. Ronzani, B. Karimi, J. Senior, Y. C. Chang, J. T. Peltonen, C. D. Chen, and J. P. Pekola, Nat. Phys. \textbf{14}, 991 (2018).
\bibitem{jsenior2020cp}J. Senior, A. Gubaydullin, B. Karimi, J. T. Peltonen, J. Ankerhold, and J. P. Pekola,Communication Physics \textbf{3}, 40 (2020).
\bibitem{dwang2019np}D. W. Wang, C. Song, W. Feng, H. Cai, D. Xu, H. Deng, H. K. Li, D. N. Zheng, X. B. Zhu, H. Wang, S. Y. Zhu, and M. O. Scully,
Nat. Phys. \textbf{15}, 382 (2019).

\bibitem{hxu2016nature}H. T. Xu, D. Mason, L. Y. Jiang, and J. G. E. Harris, Nature \textbf{537}, 80 (2016).
\bibitem{sbarik2018sci}S. Barik, A. Karasahin, C. Flower, T. Cai, H. Miyake, W. DeGottardi, M. Hafezi, and E. Waks, Science \textbf{359}, 666 (2018).
\bibitem{ywang2020pr}Y. J. Wang, J. Ren, W. X. Zhang, L. He, and X. D. Zhang, Photon. Res. \textbf{8}, B39 (2020).




\bibitem{uweiss2012book}U. Weiss, \emph{Quantum dissipative systems} (World Scientific, Singapore, 2008).

\bibitem{dsegal2006prb}D. Segal, Phys. Rev. B \textbf{73}, 205415 (2006).
\bibitem{dsegal2011prb}T. Chen, X. B. Wang, and J. Ren, Phys. Rev. B \textbf{87}, 144303 (2013).

\bibitem{ksaito2013prl}K. Saito and T. Kato, Phys. Rev. Lett. \textbf{111}, 214301 (2013).
\bibitem{mcarrega2016prl}M. Carrega, P. Solinas, M. Sassetti, and U. Weiss, Phys. Rev. Lett. \textbf{116}, 240403 (2016).
\bibitem{wdou2018prb}W. J. Dou, M. A. Ochoa, A. Nitzan, and J. E. Subotnik, Phys. Rev. B \textbf{98}, 134306 (2018).
\bibitem{anazir2020prl}H. Maguire, J. lles-Smith, and A. Nazir Phys. Rev. Lett. \textbf{123}, 093601 (2019).

\bibitem{aleggett1987rmp}A. J. Leggett, S. Chakravarty, A. T. Dorsey, Matthew P. A. Fisher, A. Garg, and W. Zwerger,
Rev. Mod. Phys. \textbf{59}, 1 (1987).

\bibitem{dsegal2005prl}D. Segal and A. Nitzan, Phys. Rev. Lett. \textbf{94}, 034301 (2005).
\bibitem{cwang2015sr}C. Wang, J. Ren, and J. S. Cao, Sci. Rep. \textbf{5} 11878 (2015).
\bibitem{cwang2017pra}C. Wang, J. Ren, and J. S. Cao, Phys. Rev. A \textbf{95} 023610 (2017).
\bibitem{xfcao2021prb}X. F. Cao, C. Wang, H. Zheng, and D. H. He, Phys. Rev. B \textbf{103}, 075407 (2021).

\bibitem{yy2014epl}Y. Yang and C. Q. Wu, Euro. Phys. Lett. \textbf{107}, 30003 (2014).
\bibitem{jjliu2017pre}J. J. Liu, H. Xu, B. Li, and C. Q. Wu, Phys. Rev. E \textbf{96}, 012135 (2017).

\bibitem{mwallquist2009ps}M. Wallquist, K. Hammerer, P. Rabl, M. Liukin, and P. Zoller, Phys. Scr. \textbf{T137}, 014001 (2009).
\bibitem{gkurizki2015pnas}G. Kurizki, P. Bertet, Y. Kubo, K. Molmer, D. Petrosyan, P. Rabl, and J. Schmiedmayer, PNAS \textbf{112}, 3866 (2015)
\bibitem{aclerk2020np}A. A. Clerk, K. W. Lehnert, P. Bertet, J. R. Petta, and Y. Nakamura, Nat. Phys. \textbf{16}, 257 (2020).

\bibitem{akockum2019nrp}A. F. Kockum, A. Miranowicz, S. De Liberato, S. Savasta, and F. Nori, Nat. Rev. Phys. \textbf{1}, 19 (2019).
\bibitem{pdiaz2019rmp}P. Forn-D\'{i}az, L. Lamata, E. Rico, J. Kono, and E. Solano, Rev. Mod. Phys. \textbf{91}, 025005 (2019).
\bibitem{maspelmeyer2014rmp} M. Aspelmeyer, T. J. Kippenberg, and F. Marquardt, Rev. Mod. Phys. \textbf{86}, 1391 (2014).

\bibitem{anazir2014pra}J. lles-Smith, N. Lambert, and A. Nazir, Phys. Rev. A \textbf{90}, 032114 (2014).
\bibitem{anazir2016jcp}J. lles-Smith, A. G. Dijkstra, N. Lambert, and A. Nazir, J. Chem. Phys. \textbf{144}, 044110 (2016).
\bibitem{dnewman2017pre}D. Newman, F. Mintert, and A. Nazir, Phys. Rev. E \textbf{95}, 032139 (2017).
\bibitem{srestrepo2018njp}S. Restrepo, J. Cerrillo, P. Strasberg, and G. Schaller, New J. Phys. \textbf{20}, 053063 (2018).
\bibitem{cmc2019jcp}C. McConnell and A. Nazir, J. Chem. Phys. \textbf{151}, 054104 (2019).

\bibitem{sfelicetti2018pra1}S. Felicetti, D. Z. Rossatto, E. Rico, E. Solano, and P. Forn-D\'{i}az, Phys. Rev. A \textbf{97}, 013851 (2018).
\bibitem{sfelicetti2018pra2}S. Felicetti, M. J. Hwang, and A. LeBoit\'{e}, Phys. Rev. A \textbf{98}, 053859 (2018).

\bibitem{qhchen2012pra}Q. H. Chen, C. Wang, S. He, T. Liu, and K. L. Wang, Phys. Rev. A \textbf{86}, 023822 (2012).
\bibitem{lduan2016jpa}L. W. Duan, Y. F. Xie, D. Braak, and Q. H. Chen, J. Phys. A \textbf{49}, 464002 (2016).

\bibitem{lgarbe2017pra}L. Garbe, I. L. Egusquiza, E. Solano, C. Ciuti, T. Coudreau, P. Milman, and S. Felicetti, Phys. Rev. A \textbf{95}, 053854 (2017).
\bibitem{xchen2018pra}X. Y. Chen and Y. Y. Zhang, Phys. Rev. A \textbf{97}, 053821 (2018).
\bibitem{scui2020pra}S. F. Cui, B. Gr\'{e}maud, W. A. Guo, and G. G. Batrouni, Phys. Rev. A \textbf{102}, 033334 (2020).

\bibitem{ac2020prb}A. Crescente, M. Carrega, M. Sassetti, and D. Ferraro, Phys. Rev. B \textbf{102}, 245407 (2020).
\bibitem{ad2021entropy}A. Delmonte, A. Crescente, M. Carrega, D. Ferraro, and M. Sassetti, Entropy \textbf{23}, 612 (2021).
\bibitem{nparxiv2020}N. Piccione, S. Felicetti, and B. Bellomo, arXiv:2007.07844.
\bibitem{jjliu2020nl}J. J. Liu and D. Segal, Nano. Lett. \textbf{20}, 6128 (2020).

\bibitem{aseif2017nc}A. Seif, W. DeGottardi, K. Esfarjani, and M. Hafezi, Nat. Comm. \textbf{9}, 1207 (2017).
\bibitem{zdenis2020prl}Z. Denis, A. Biella, I. Favero, and C. Ciuti, Phys. Rev. Lett. \textbf{124}, 083601 (2020).
\bibitem{wjnie2020pra}W. J. Nie, G. Y. Li, X. Y.  Li, A. X. Chen, Y. H. Lan, and S. Y. Zhu, Phys. Rev. A \textbf{102}, 043512 (2020).

\bibitem{yang2020nc}C. Yang, X. R. Wei, J. T. Sheng, and H. B. Wu, Nat. Comm. \textbf{11}, 4656 (2020).

\bibitem{mmajland2020prb}M. Majland, K. S. Christensen, and N. T. Zinner, Phys. Rev. B \textbf{101}, 184510 (2020).
\bibitem{cwang2021cpl}C. Wang, L. Q. Wang, and J. Ren, Chin. Phys. Lett. \textbf{38}, 010501 (2021).
\bibitem{cwang2021cpb}C. Wang, L. Q. Wang, and J. Ren, Chin. Phys. B \textbf{30}, 030506 (2021).

\bibitem{yjzhao2015pra} Y. J. Zhao, Y. L. Liu, Y. X. Liu, and F. Nori, Phys. Rev. A \textbf{91}, 053820 (2015).
\bibitem{xwang2016pra} X. Wang, A. Miranowicz, H. R. Li, and F. Nori, Phys. Rev. A \textbf{94}, 053858 (2016).
\bibitem{xwang2017pra} X. Wang, A. Miranowicz, H. R. Li, and F. Nori, Phys. Rev. A \textbf{96}, 063820 (2017).
\bibitem{sricher2016prb} S. Richer and D. Divincenzo, Phys. Rev. B \textbf{93}, 134501 (2016).
\bibitem{nlambert2018prb} N. Lambert, M. Cirio, M. Delbecq, G. Allison, M. Marx, S. Tarucha, and F. Nori, Phys. Rev. B \textbf{97}, 125429 (2018).


\bibitem{tkato2020arxiv}T. Yamamoto and T. Kato, arXiv:2011.06815.

\bibitem{aaclerk2010rmp} A. A. Clerk, M. H. Devoret, S. M. Girvin, F. Marquardt, and R. J. Schoelkopf, Rev. Mod. Phys. \textbf{82}, 1155 (2010).


\bibitem{asettineri2018pra}A. Settineri, V. Macr\'{i}, A. Ridolfo, O. Di Stefano, A. F. Kockum, F. Nori, and S. Savasta, Phys. Rev. A \textbf{98}, 053834 (2018).


\bibitem{pkral1990jmo}P. Kral, J. Mod. Opt. \textbf{37}, 889 (1990)


\bibitem{levitov1992jetp}L. Levitov and G. Lesovik, JETP Lett. \textbf{55}, 555 (1992).
\bibitem{levitov1996jmp}L. Levitov, H. Lee, and G. Lesovik, J. Math. Phys. \textbf{37}, 4845 (1996).

\bibitem{mesp2009rmp}M. Esposito, U. Harbola, and S. Mukamel, Rev. Mod. Phys. \textbf{81}, 1665 (2009).


\bibitem{bli2006apl}B. Li, L. Wang, and G. Casati, Appl. Phys. Lett. \textbf{88}, 143501 (2006).
\bibitem{nbli2012rmp}N. B. Li, J. Ren, L. Wang, G. Zhang, P. Hanggi, and B. Li, Rev. Mod. Phys. \textbf{84}, 1045 (2012).

\bibitem{jren2013prb}J. Ren and J. X. Zhu, Phys. Rev. B \textbf{87}, 241412(R) (2013).
\bibitem{afornieri2016prb}A. Fornieri, G. Timossi, R. Bosisio, P. Solinas, and F. Giazotto, Phys. Rev. B \textbf{93}, 134508 (2016).


\bibitem{bli2004prl}B. Li, L. Wang, and G. Casati, Phys. Rev. Lett. \textbf{93}, 184301 (2004).
\bibitem{mjmperez2015nn}M. J. M. P\'{e}rez, A. Fornieri, and F. Giazotto, Nat. Nanotech. \textbf{10}, 303 (2015).

\bibitem{lfzhang2009prb}L. F. Zhang, Y. H. Yan, C. Q. Wu, J. S. Wang, and B. Li, Phys. Rev. B \textbf{80}, 172301 (2009).

\bibitem{fzhan2011prb}F. Zhan, S. Denisov, and P. H\"{a}nggi, Phys. Rev. B \textbf{84}, 195117 (2011).
\bibitem{rbelousov2016pre}R. Belousov, E. G. D. Cohen, C. S. Wong, J. A. Goree, and Y. Feng, Phys. Rev. E \textbf{93}, 042125 (2016).

\bibitem{nasinitsyn2007epl}N. A. Sinitsyn and I. Nemenman, Europhys. Lett. \textbf{77}, 58001 (2007).
\bibitem{jren2010prl}J. Ren, P. H\"{a}nggi, and B. Li, Phys. Rev. Lett. \textbf{104}, 170601 (2010).






\bibitem{dstoler1970prd}D. Stoler, Phys. Rev. D \textbf{1}, 3217 (1970).
\bibitem{dstoler1971prd}D. Stoler, Phys. Rev. D \textbf{4}, 1925 (1971).
\bibitem{ula2016ps}U. L. Andersen, T. Gehring, C. Marquardt, and G. Leuchs, Phys. Scr. \textbf{91}, 053001 (2016).

\bibitem{jrk1988pra} J. R. Kukli\'{n}ski and J. L. Madajczyk, Phys. Rev. A \textbf{37}, 3175 (1988).
\bibitem{ax2020np} A. Xuereb, Nat. Phys. \textbf{16}, 707 (2020).

\bibitem{as2013prl} A. Szorkovszky, G. A. Brawley, A. C. Doherty, and W. P. Bowen, Phys. Rev. Lett. \textbf{110}, 184301 (2013).
\bibitem{ak2013pra} A. Kronwald, F. Marquardt, and A. A. Clerk, Phys. Rev. A \textbf{88}, 063833 (2013).
\bibitem{eew2015science} E. E. Wollman, C. U. Lei, A. J. Weinstein, J. Suh, A. Kronwald, F. Marquardt, A. A. Clerk, and K. C. Schwab,
Science \textbf{349}, 952 (2015).

\bibitem{jma2011pr}J. Ma, X. G. Wang, C. P. Sun, and F. Nori, Phys. Rep. \textbf{509}, 89 (2011).


\bibitem{jqliao2011pra} J. Q. Liao and C. K. Law, Phys. Rev. A \textbf{83}, 033820 (2011).
\bibitem{xylu2015pra} X. Y. L\"{u}, J. Q. Liao, L. Tian, and F. Nori, Phys. Rev. A \textbf{91}, 013834 (2015).

\bibitem{mm2008prl}M. Marthaler, G. Sch\"{o}n, and A. Shnirman, Phys. Rev. Lett. \textbf{101}, 147001 (2008).
\bibitem{jkxie2020pra} J. K. Xie, S. L. Ma, Y. L. Ren, X. K. Li, and F. L. Li, Phys. Rev. A \textbf{101}, 012348 (2020).




\end{thebibliography}
\end{document}